\documentclass[aps,prl,reprint,superscriptaddress,amsmath]{revtex4-1}

\makeatletter

\usepackage{changes}
\usepackage{graphicx}
\usepackage{dcolumn}
\usepackage{bm}
\usepackage[version=4]{mhchem}
\usepackage{graphicx}
\usepackage{color}     
\newcommand{\RNum}[1]{\uppercase\expandafter{\romannumeral #1\relax}}

\makeatother

\begin{document}
\title{Crystalline Electric-Field Excitations in Quantum Spin Liquids Candidate \ce{NaYbSe2}}

\author{Zheng\,Zhang}
\affiliation{Beijing National Laboratory for Condensed Matter Physics, Institute of Physics, Chinese Academy of Sciences, Beijing 100190, China}
\affiliation{Department of Physics, Renmin University of China, Beijing 100872, China}

\author{Xiaoli\,Ma}
\affiliation{Beijing National Laboratory for Condensed Matter Physics, Institute of Physics, Chinese Academy of Sciences, Beijing 100190, China}

\author{Jianshu\,Li}
\affiliation{Beijing National Laboratory for Condensed Matter Physics, Institute of Physics, Chinese Academy of Sciences, Beijing 100190, China}
\affiliation{Department of Physics, Renmin University of China, Beijing 100872, China}

\author{Guohua\,Wang}
\affiliation{Department of Physics and Astronomy, Shanghai Jiao Tong University, Shanghai 200240, China}

\author{D.T.\,Adroja}
\affiliation{ISIS Neutron and Muon Facility, SCFT Rutherford Appleton Laboratory, Chilton, Didcot Oxon, OX11 0QX, United Kingdom}
\affiliation{Highly Correlated Matter Research Group, Physics Department, University of Johannesburg, Auckland Park 2006, South Africa}

\author{T.G.\,Perring}
\affiliation{ISIS Neutron and Muon Facility, SCFT Rutherford Appleton Laboratory, Chilton, Didcot Oxon, OX11 0QX, United Kingdom}

\author{Weiwei\,Liu}
\affiliation{Beijing National Laboratory for Condensed Matter Physics, Institute of Physics, Chinese Academy of Sciences, Beijing 100190, China}
\affiliation{Department of Physics, Renmin University of China, Beijing 100872, China}

\author{Feng\,Jin}
\affiliation{Beijing National Laboratory for Condensed Matter Physics, Institute of Physics, Chinese Academy of Sciences, Beijing 100190, China}

\author{Jianting\,Ji}
\affiliation{Beijing National Laboratory for Condensed Matter Physics, Institute of Physics, Chinese Academy of Sciences, Beijing 100190, China}

\author{Yimeng\,Wang}
\affiliation{Beijing National Laboratory for Condensed Matter Physics, Institute of Physics, Chinese Academy of Sciences, Beijing 100190, China}
\affiliation{Department of Physics, Renmin University of China, Beijing 100872, China}

\author{Xiaoqun\,Wang}
\affiliation{Department of Physics and Astronomy, Shanghai Jiao Tong University, Shanghai 200240, China}

\author{Jie\,Ma}
\email[e-mail:]{jma3@sjtu.edu.cn}
\affiliation{Department of Physics and Astronomy, Shanghai Jiao Tong University, Shanghai 200240, China}

\author{Qingming\,Zhang}
\email[e-mail:]{qmzhang@ruc.edu.cn}
\affiliation{School of Physical Science and Technology, Lanzhou University, Lanzhou 730000, China}
\affiliation{Beijing National Laboratory for Condensed Matter Physics, Institute of Physics, Chinese Academy of Sciences, Beijing 100190, China}

\begin{abstract}
Very recently we revealed a large family of triangular lattice quantum spin liquid candidates named rare-earth chalcogenides, which features a high-symmetry structure without structural/charge disorders and spin impurities, and may serve as an ideal platform exploring spin liquid physics.  The knowledge of crystalline electric-field (CEF) excitations is an essential step to explore the fundamental magnetism of rare-earth spin systems.
Here we employed inelastic neutron scattering (INS) and Raman scattering (RS) to carry out a comprehensive CFE investigation on \ce{NaYbSe2},  a promising representative of the family. By comparison with its nonmagnetic compound \ce{NaLuSe2}, we are able to identify the CEF excitations at 15.8, 24.3 and 30.5 meV at 5K. The selected cuts of the INS spectra are well re-produced with a large anisotropy of $g$ factors ($g_{ab}:g_{c}\sim3:1$). Further, the CEF excitations are explained well by our calculations based on the point charge model. Interestingly, \ce{NaYbSe2} exhibits an unusual CEF shift to higher energies with increasing temperatures, and the Raman mode close to the first CEF excitation shows an anomalously large softening with decreasing temperatures. The absence of the anomalies in \ce{NaLuSe2} clearly demonstrates a CEF-phonon coupling not reported in the family. It can be understood in term of the weaker electronegativity of \ce{Se}. The fact that the smallest first CEF excitation in the sub-family of \ce{NaYbCh2} is $\sim$ 180K (Ch=O, S, Se), guarantees that the sub-family can be strictly described with an effective S=1/2 picture at sufficiently low temperatures. Interestingly the CEF-phonon coupling revealed here may present alternative possibilities to manipulate the spin systems. 
\end{abstract}

\maketitle
\section{Introduction}
\begin{figure*}
	\includegraphics[scale=0.65]{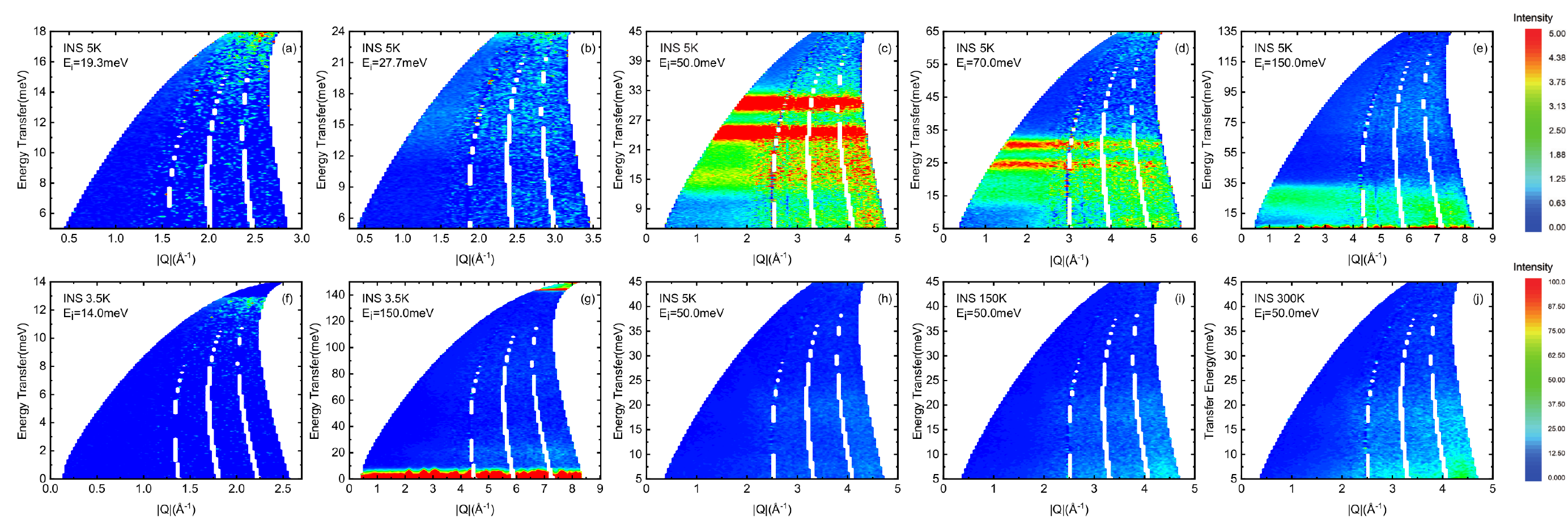}
	\caption{\label{fig:wide}Upper panel: INS spectra of \ce{NaYbSe2} measured with different incident neutron energies at 5K. Lower panel: INS spectra of non-magnetic \ce{NaLuSe2} for comparison.}
\end{figure*}

Quantum spin liquid (QSL) is a novel quantum spin-entangled state breaking no symmetry even at zero temperature\cite{Balents2010,Savary2016}. The idea of QSL was proposed by P.W Anderson in 1973 and then applied to high temperature superconductivity\cite{Anderson1973,ANDERSON1987}. A lot of theoretical effort has been contributed to the exotic state since then. X.G.Wen made a topological classification for QLSs on the mean-field level\cite{Wen1991}. In 2006, Kitaev proposed an exactly solvable model on a spin-1/2 honeycomb lattice, which inspires topological quantum computing based on strong spin-entanglement of QSL\cite{Kitaev2006}.

Meanwhile, much effort has also been made on the material side. Herbersmithite is a famous QSL candidate with 3d spin-1/2 \ce{Cu^{2+}} ions sitting on the kagome lattice\cite{Shores2005,Han2012}, while the anti-site mixing between \ce{Cu^{2+}} and \ce{Zn^{2+}} ions remains a key issue in uncovering the intrinsic properties of the spin ground state. We turned to strong spin-orbital (SO) coupling and discovered the rare-earth-based QSL candidate \ce{YbMgGaO4}, which possesses a perfect spin triangular lattice and and rules out the anti-site mixing, spin impurities and Dyaloshinski-Moriya interaction \cite{Li2015a,Li2015}. A series of experiments down to milikelvins, such as thermodynamic measurements, muon spin relaxation (muSR) and INS etc., consistently suggest a gapless QSL ground state\cite{Li2015,Li2016a,Li2017,Shen2016,Li2019}. On the other hand, it was concerned that the charge disorder between \ce{Mg}/\ce{Ga} in \ce{YbMgGaO4}\cite{LiYuSheng_CEF} may have influence on the spin channel.

The idea of strong SO coupling continuously leads us to reveal a large family of rare-earth-based QSL candidates \ce{ARECh2} (A=Alkali or monovalent ion, RE=rare earth, Ch=chalcogen)\cite{Liu2018}. The family shares the same high symmetry with \ce{YbMgGaO4} and inherits almost all the advantages of it but naturally removes the issue of charge disorder. Most excitingly, the family has a huge diversity of members and the preliminary results show that the antiferromagnetic (AF) exchange coupling is improved by at least one order of magnitude compared to \ce{YbMgGaO4}. These enable the family as a unique and ideal playground for spin liquid study. 

The exploration on the sub-family of \ce{NaYbCh2} (Ch=chalcogen) has been launched\cite{Liu2018,Baenitz2018,Ranjith2019,Ding2019a,Sichelschmidt2019,Ranjith2019a,Xing2019,Bordelon2019,Xing2019a,Xing2019b,Gao2019,Sichelschmidt2019a,Sarkar2019}. The knowledge of the CFE excitations is an essential step to look into the fundamental magnetism of the family. Among the \ce{NaYbCh2} members, \ce{NaYbSe2} is of particular interest since it shows a distinct QSL signature and an appropriate charge gap ($\sim$ 1.92 eV) which is crucial to the metallization and material engineering\cite{Liu2018}. Another interesting issue is the essential influence of the ligand anions on structural and magnetic properties, for which we expect to find some key clues by investigating the CEF excitations of \ce{NaYbSe2}. At present we are able to grow \ce{NaYbSe2} single crystals with the largest dimension ($\sim$ 20mm) among the sub-family members. This makes single-crystal neutron scattering experiments possible and it is highly required to figure out the CEF excitations of \ce{NaYbSe2} before inelastic neutron scattering experiments.

In this paper, by applying INS and RS to \ce{NaYbSe2} and its nonmagnetic reference compound \ce{NaLuSe2}, we made a comprehensive investigation on the CEF excitations in \ce{NaYbSe2} that revealed an unusual CEF-Phonon coupling. Three CEF excitations, at 15.79meV, 24.33meV and 30.53meV, are probed at 5K by INS experiments, and further confirmed by RS experiments. Our simulations well re-produce the selected cuts of the INS spectra and we employed the point charge model to calculate the CEF excitations which are consistent with the experimental ones. Both INS and RS experiments demonstrate that the CEF excitations unexpectedly shift to higher energies with increasing temperatures. Moreover, the Raman phonon mode very close to the first CEF excitation is significantly softened with decreasing temperatures and its width also shows an abnormal behavior, while the anomalies completely disappear in \ce{NaLuSe2}. These allow to identify a CEF-phonon coupling in \ce{NaYbSe2}.  The phonon anomalies as well as the unusual CEF temperature dependence, can be explained in term of the electronegativity of chalcogen elements.

\section{Experimental Techniques}

The high-quality single crystals of \ce{NaYbSe2}, as well as the nonmagnetic \ce{NaLuSe2}, were grown using a \ce{NaCl}-flux method. The single crystals with the dimensions of 5mm $\times$ 5mm were used in RS experiments. The polycrystalline \ce{NaYbSe2} and \ce{NaLuSe2} were synthesized by the high-temperature solid state reaction and were characterized to be single-phased\cite{Liu2018}. $\sim$5g powder sample of \ce{NaYbSe2} and $\sim$3g powder sample of \ce{NaLuSe2} were used in INS experiments (See supplementary materials for details). 
The INS data on \ce{NaYbSe2} and nonmagnetic reference compound \ce{NaLuSe2} were collected using the high flux and high resolution time of flight spectrometers, MAPS, at the ISIS pulsed neutron facility, Rutherford Appleton Laboratory, U.K.  The energy- and temperature-dependence of data  were obtained. The powder samples were loaded in a cylinder Al-can with an inner diameter of 30mm. The closed cycle refrigerator (CCR) was used to cool the samples to a based temperature of 5K with He-exchange gas. 
The Raman spectra were collected using a HR800 (Jobin Yvon) and T64000 (Jobin Yvon) equipped with a 633nm and 473nm laser, charge-coupled device(CCD) and volume Bragg gratings. After cleavage, the single crystals of \ce{NaYbSe2} and \ce{NaLuSe2} were placed in a closed-cycle cryostat for Raman experiments. Backscattering configuration was employed and the polarizations of both incident and scattering light lie in the ab plane.

\section{CEF excitations and Inelastic neutron scattering}

\begin{figure*}
	\includegraphics[scale=0.7]{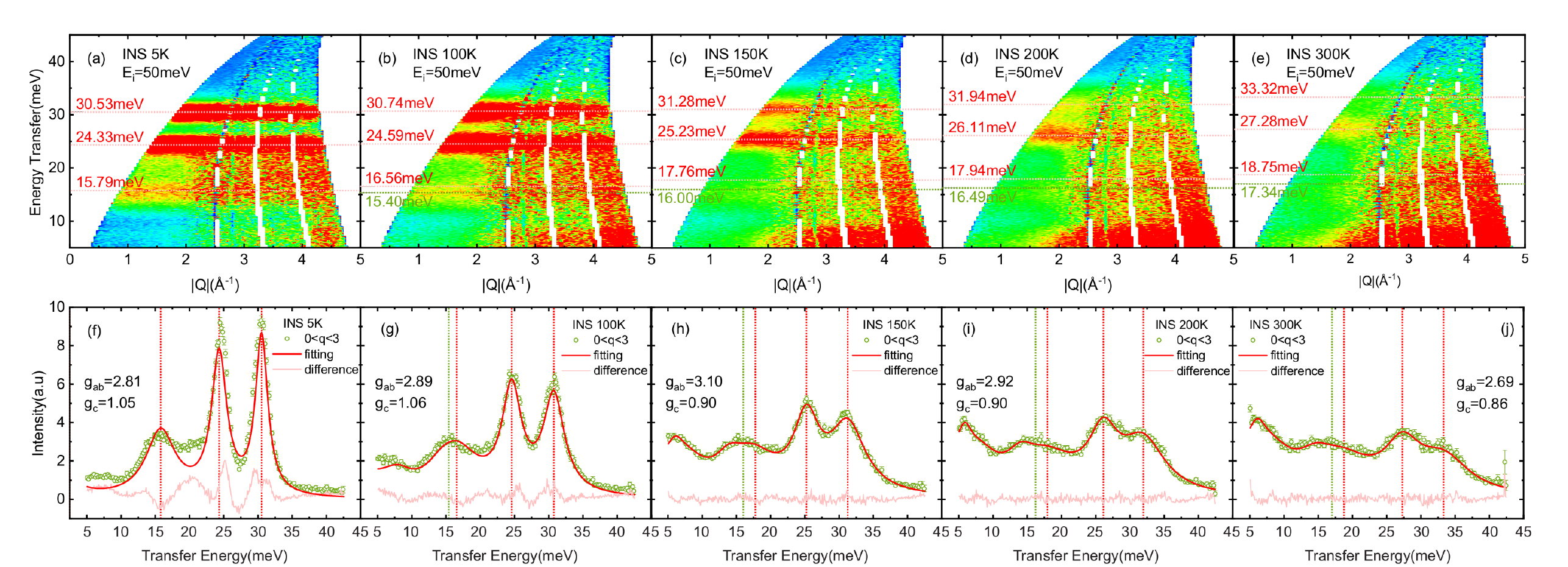}
	\caption{\label{fig:wide} (a)$\sim$(e) are the INS spectra for \ce{NaYbSe2} measured with the incident neutron energy $E_{i}=50$ meV at selected temperatures. (f)$\sim$(j) are the cuts of INS intensity versus energy corresponding to  (a)$\sim$(e). The data in (f)$\sim$(j) are obtained by integrating over the wavevector space from 0 to 3 \AA$^{-1}$ (green dots). The fitting curves (red lines) are given by multiple iterations with initial CEF parameters calculated from the point charge mode. In (a)$\sim$(e) and (f)$\sim$(j), the fitted 1st, 2nd and 3rd CEF excitations of \ce{Yb^{3+}} are highlighted by dashed red lines. The dashed green lines above 100 K mark $E_g$ phonon mode which is identified in Raman experiments.}
\end{figure*}

The INS spectra for \ce{NaYbSe2} with the incident neutron energy $E_{i}=$19.3, 27.7, 50.0, 70.0 and 150.0 meV at 5K are shown in Fig. 1 (upper panel). Three excitations can be clearly seen at 15.8, 24.3 and 30.5 meV, respectively. For comparison, the INS spectra of non-magnetic reference sample \ce{NaLuSe2} are also shown here (lower panel of Fig. 1) and only weak phonon excitations are observed. We have the similar observations in Raman experiments. In Raman channel, the excitations are observed in \ce{NaYbSe2} while they are absent in \ce{NaLuSe2}. The observed excitations exhibit the moment-independence and their absence in \ce{NaLuSe2}, which rule out the phononic origin of them, and hence can be unambiguously identified as the CEF ones.

It is interesting that the first CEF excitation of 15 meV($\sim$ 185K) is smaller than that of \ce{YbMgGaO4} (39meV, $\sim$ 480K)\cite{LiYuSheng_CEF} and that of \ce{NaYbO2} (36meV, $\sim$ 440K)\cite{Ding2019a}. It suggests that the larger Se-Yb bond length and the weaker electronegativity of selenium effectively reduce the CEFs at Yb ions. The first CEF excitation sets a fundamental energy scale for studying the magnetism of \ce{NaYbSe2} and allows to make a relatively accurate Curie-Weiss analysis.

Fig. 2 demonstrates the temperature dependence of the CEF excitations. The cuts of the integrated spectra in the ranges of low wavevector (0 $\sim$ 3 \AA$^{-1}$) and high wavevector (3 $\sim$ 5 \AA$^{-1}$) are shown in Fig.2(f) $\sim$ 2(j). There are several points we can draw from the spectra.

(\romannumeral1) The intensity of the three CEF excitations decrease with $|Q|$. This can be explained by the magnetic form factor $F(|Q|)$ related to magnetic ions in the differential neutron scattering cross section. We calculated the magnetic form factor $F^{2}(|Q|)$ as a function of $|Q|$, and found that the scattering intensity decreases with increasing $|Q|$, consistent with the observations here (See supplementary material).

(\romannumeral2) The intensity of the CEF excitations decreases with increasing temperature. This simply comes from the population factor and thermal broadening and the similar behavior has been also seen in \ce{YbMgGaO4}\cite{LiYuSheng_CEF} and \ce{NaYbS2}\cite{Baenitz2018}.

(\romannumeral3) Most interestingly, the CEF excitations exhibit a slight but clear shift to higher energies with increasing temperatures. Our Raman experiments also catch the shift (Fig. 4). This unusual temperature dependence was not observed in \ce{YbMgGaO4}\cite{LiYuSheng_CEF} and \ce{NaYbS2}\cite{Baenitz2018}. The temperature-dependent CEF excitations are quite unusual. This seems related to the larger radius of \ce{Se} anion or its weaker electronegativity. At lower temperatures, there is a larger overlap of electron cloud between \ce{Yb} and \ce{Se}, which reduces the effective charges of \ce{Se} anions and hence gives rise to the lower CEF levels based on the point charge model. With increasing temperatures, the general lattice expansion slightly but continuously enlarges the bond distance between \ce{Yb} and \ce{Se} and reduces the electron cloud overlap between them. This slightly enhances the ionicity of the compound and raises the effective charge of \ce{Se}, and consequently lifts the CEF levels in a small amount. The thermal broadening of the CEF levels at higher temperatures actually makes the lift obscure, as observed in neutron experiments. 

(\romannumeral4) The first CEF energy level above 100 K deviates from the fitted one (FIG. 2(f) $\sim$ 2(j)). This may be related to the coupling between CEF excitations and phonons. The CEF excitations move towards high energy with increasing temperatures, while the coupling enables an energy exchange between the first CEF excitation and phonons which are very close in energy (See below). Eventually the first CEF excitation exhibits a smaller shift with temperature compared to the second and third excitations. The effect is also seen by RS experiments (See below). In the reference sample \ce{NaLuSe2}, only one broad peak has been observed and the its position decreases with increasing temperatures(See supplementalary material). It can be identified as a phonon mode through our RS experiments.

We have determined the CEF parameters by carefully fitting the cuts of INS spectra in the range from $0$ to $3$ \AA$^{-1}$. The fitting details and the the extracted CEF parameters can be found in the supplementary materials.The CEF parameters can be calculated by using formulas (See supplementary material). The calculated g-factors are shown in the panel of Fig. 2(f) $\sim$ 2(j). The ratio of $g_{ab-plane}$ to $g_{c-axis}$ is $\sim$ 3:1, close to that in \ce{NaYbO2} and \ce{NaYbS2}\cite{Ranjith2019,Baenitz2018}. This indicates that the sub-family \ce{AYbCh2} has a systematic and strong magnetic anisotropy, which is much less in \ce{YbMgGaO4} where the ratio is close to 1\cite{Li2015a}. The difference may stem from the charge imbalance of different cations between \ce{Yb^{3+}} layers\cite{Zangeneh2019}.

\begin{figure}[b]
	\includegraphics[scale=0.7]{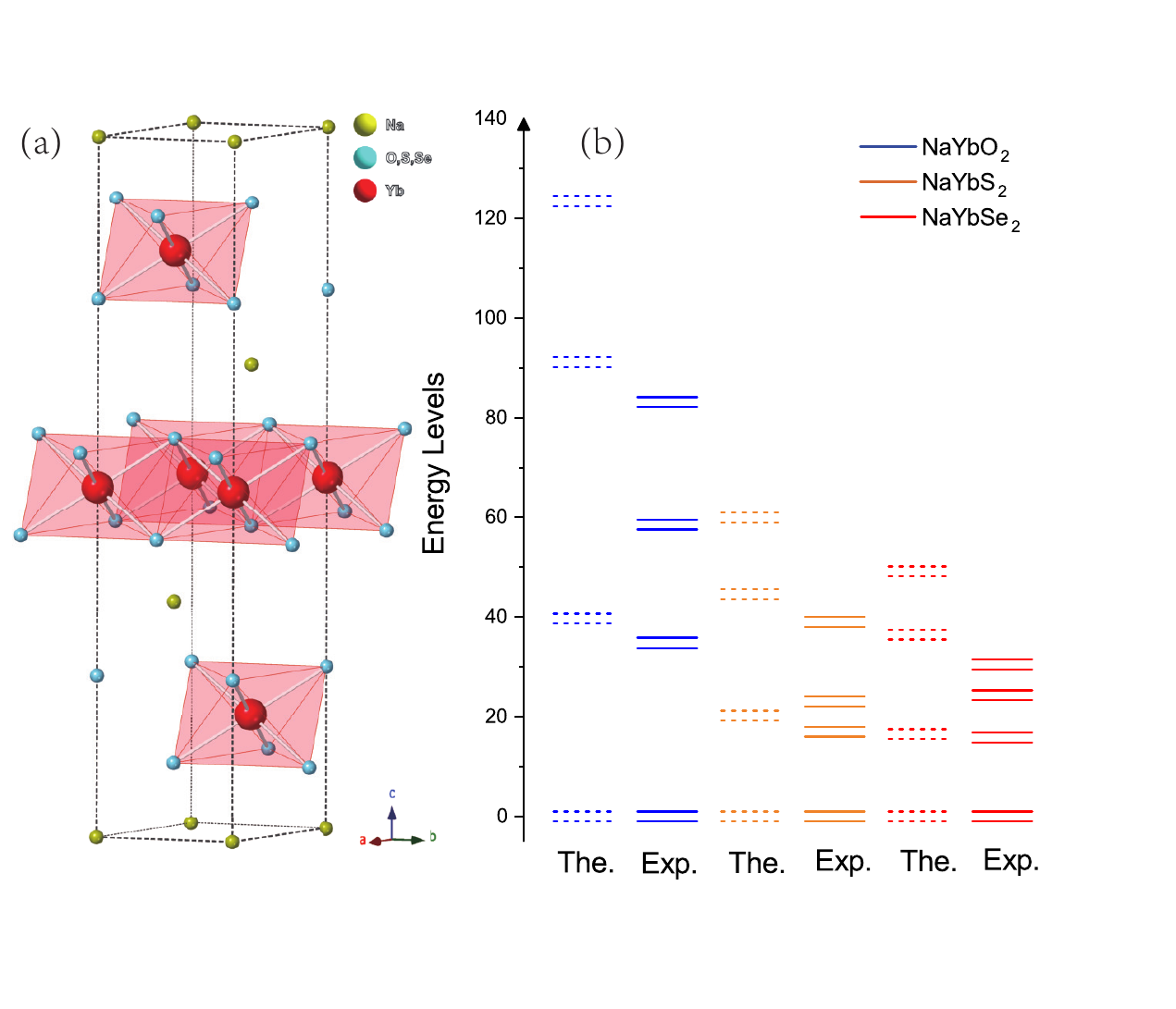}
	\caption{\label{fig:epsart} (a) Crystal structure of \ce{AYbCh2}(Ch=O, S, Se). (b) Experimental CEF excitation energy levels extracted from INS experiments (solid lines) and the calculated ones based on the point charge model (dashed lines).}
\end{figure}

We have calculated the CEF levels in \ce{NaYbO2}, \ce{NaYbS2} and \ce{NaYbSe2} using the point charge model (See supplementary materials), and made comparison with the experimental ones (Fig. 3). Generally the CEF levels decrease with the radii of \ce{O},\ce{S} and \ce{Se} as expected. One can see a good agreement between experiments and calculations for the first CEF level. The calculated first CEF levels  are $\sim$39.7 meV (\ce{NaYbO2}), $\sim$20.2 meV (\ce{NaYbS2}), and $\sim$ 16.52 meV (\ce{NaYbSe2}), and the experimental values obtained from INS measurements are 34.8meV \cite{Ding2019a}, 17 meV\cite{Baenitz2018}, and 15.8 meV, respectively. It implies that the ionic crystal picture still works well even for the case with larger selenium anions. When we look at the second and third excitation levels, there is a relatively large discrepancy between the experimental and calculated ones. Reducing the discrepancy actually requires a higher-order and accurate description on the CEF environments. The results also suggest that though the magnetism in the QSL candidates is dominated by rare-earth magnetic ions, the surrounding coordination anions(\ce{O}, \ce{S} and \ce{Se} etc.) and the cations between magnetic planes have impact on spin states.

\section{CEF-Phonon coupling and Raman scattering}
\begin{figure*}
	\includegraphics[scale=0.3]{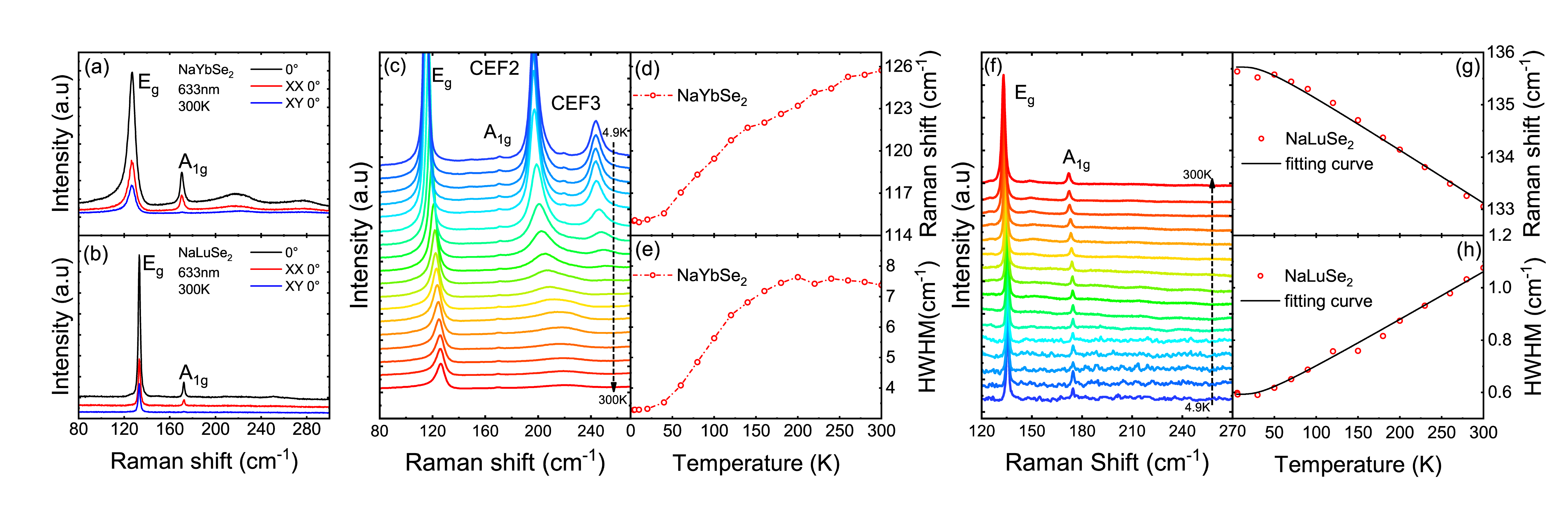}
	\caption{\label{fig:wide} Polarized Raman spectra for \ce{NaYbSe2} (a) and \ce{NaLuSe2} (b). XX and XY represent parallel and cross polarization configurations, respectively. (c) and (f) are the temperature-dependent Raman spectra of \ce{NaYbSe2} and \ce{NaLuSe2}, respectively, and the Raman shift and half widths at half-maximum (HWHM) of $E_g$ mode are shown in panle (d), panle (e) and panle (g), panle (h).}
\end{figure*}

Fig. 4 shows Raman spectra of \ce{NaYbSe2} and \ce{NaLuSe2}. The crystal symmetry $R-3m$ allows two Raman-active phonon modes $A_{g}+E_{g}$. Symmetry analysis indicates that $[A_{g}]$ mode is visible only in the parallel polarization configuration while $E_{g}$ mode can be observed for in both parallel and cross polarization configurations. The two modes can be clearly identified with the polarized spectra in Fig. 4(a) and 4(b).

Besides the Raman modes, two weaker bands appear at 217 and 277 cm$^{-1}$ in the spectra of \ce{NaYbSe2}. The frequencies exactly match the 2nd and 3rd CEF levels given by INS measurements. And the bands completely disappear in \ce{NaLuSe2}. Therefore, the two bands can be safely assigned to the CEF excitations. In fact, Raman scattering is a unique method to probe CEF excitations and has been applied to many rare-earth-based spin systems \ce{Tb2Ti2O7}\cite{Guentherodt1983, Maczka2008, Sethi2019}. It is worth noting that the intensities of the two bands are quite different in the parallel and cross polarization configurations. This may be related to the strong anisotropy in \ce{NaYbSe2}.

There are several obvious anomalies in the Raman spectra of \ce{NaYbSe2}: 1) phonon mode broadening; 2) asymmetric lineshape of $E_{g}$ mode; 3) particularly $E_{g}$ mode shows a completely unusual softening with decreasing temperatures (Fig. 4(d)), and its width also has an abnormal temperature dependence (Fig. 4(e)). In contrast, all the anomalies are absent in the non-magnetic reference compound \ce{NaLuSe2}. This unambiguously points to a magnetic origin. Spin-phonon coupling can be ruled out, because the broadening and asymmetry of $E_{g}$ mode already exist even at room temperature which is almost one order of magnitude larger than exchange coupling. Thus we can identify that the anomalies stem from the CEF-phonon coupling. This is strongly supported by the fact that the energy scales of the CEF excitations exactly overlap with phonon frequencies. Particularly for $E_{g}$ mode, its frequency is very close to the first CEF level, though the first excitation is too weak to be observed in Raman channel perhaps due to its quite large broadening as seen in INS measurements (Fig. 2). The CEF-phonon coupling has been reported in many rare-earth compounds \cite{Guentherodt1983, Hense_2004, Sethi2019}. It may provide alternative possibility to tune the CEF excitations and the magnetism of the rare-earth spin system. 

The CEF-phonon coupling is considered to share the similar origin resulting in the unusual CEF temperature dependence as discussed above, and involve the degeneracy of the first CEF excitation and $E_{g}$ mode. The larger electron cloud overlap between \ce{Yb} and \ce{Se} in principle gives rise to a CEF-phonon coupling. The coupling in \ce{NaYbSe2} is dramatically magnified by the resonance-like effect, which originates from the fact that the first CEF excitation and $E_{g}$ mode are very close in energy. More precisely, the first CEF excitation is only $\sim$ 1 meV higher than the phonon energy at room temperature. With decreasing temperatures, the phonon frequency normally goes up while the first CEF level goes down as discussed above. The tiny energy discrepancy will be immediately filled and the resonance-like effect occurs. Generally the effect splits the resonance energy into the lower (phonon) and upper (CEF) branches. This effectively gives a phonon softening. The first CEF level seems to be dominated by the enhancement of \ce{Se} electronegativity and remains decreasing with lowering temperatures.

\section{Conclusion}

The CEF excitations of \ce{Yb^{3+}} in \ce{NaYbSe2} have been comprehensively studied using INS and Raman experiments and by comparison with non-magnetic reference sample \ce{NaLuSe2}. Three CEF excitations are observed by INS experiments and the cuts of INS spectra are well simulated to extract the CEF parameters. Based on the point charge model, we are able to reproduce the CEF excitations of \ce{NaYbO2}, \ce{NaYbS2} and \ce{NaYbSe2}, which are well consistent with the existing experimental observations. 

Despite the common characteristics of the sub-family \ce{NaYbCh2}, \ce{NaYbSe2} exhibits some unique features in the CEF excitations compared to \ce{NaYbO2} and \ce{NaYbS2}. One is that the observed CEF excitations show an interesting shift to higher energies with increasing temperatures. This is naturally understood in term of electronegativity. Another one is the obvious CEF-phonon coupling. The comparison of INS spectra of \ce{NaYbSe2} and \ce{NaLuSe2}, allow us to identify a phonon mode around 15 meV, which is very close to the first CEF excitation of \ce{NaYbSe2}. We double-checked the CEF excitations and the phonon mode in Raman channel. Surprisingly, the \ce{Eg} phonon mode close to the first CEF excitation shows a large softening with decreasing temperatures, while the same mode in \ce{NaLuSe2} has a normal hardening. The contrast is a clear demonstration of CEF-phonon coupling. The unusual CEF temperature dependence stems from the weaker electronegativity which, combined with the resonance-like process, also explains the anomalous phonon softening.

\begin{acknowledgments}
	This work was supported by the Ministry of Science and Technology of China (2017YFA0302904 \& 2016YFA0300500) and the NSF of China (11774419 \& U1932215). The data processing of INS spectrum in FIG. 1 and FIG. 2 are based on mantid software\cite{Arnold2014, Andrew2019}. The CEF energy levels calculation is based on our own MATLAB program. The theoretical calculation of CEF based on point charge model is also based on mantid software\cite{Arnold2014, Andrew2019}. We thank the Ross Stewart for help on MAPS and ISIS Facility for beam time, RB1910575. The experimental data can be found at https://doi.org/10.5286/ISIS.E.RB1910575-1
\end{acknowledgments}


\begin{thebibliography}{35}%
	\makeatletter
	\providecommand \@ifxundefined [1]{%
		\@ifx{#1\undefined}
	}%
	\providecommand \@ifnum [1]{%
		\ifnum #1\expandafter \@firstoftwo
		\else \expandafter \@secondoftwo
		\fi
	}%
	\providecommand \@ifx [1]{%
		\ifx #1\expandafter \@firstoftwo
		\else \expandafter \@secondoftwo
		\fi
	}%
	\providecommand \natexlab [1]{#1}%
	\providecommand \enquote  [1]{``#1''}%
	\providecommand \bibnamefont  [1]{#1}%
	\providecommand \bibfnamefont [1]{#1}%
	\providecommand \citenamefont [1]{#1}%
	\providecommand \href@noop [0]{\@secondoftwo}%
	\providecommand \href [0]{\begingroup \@sanitize@url \@href}%
	\providecommand \@href[1]{\@@startlink{#1}\@@href}%
	\providecommand \@@href[1]{\endgroup#1\@@endlink}%
	\providecommand \@sanitize@url [0]{\catcode `\\12\catcode `\$12\catcode
		`\&12\catcode `\#12\catcode `\^12\catcode `\_12\catcode `\%12\relax}%
	\providecommand \@@startlink[1]{}%
	\providecommand \@@endlink[0]{}%
	\providecommand \url  [0]{\begingroup\@sanitize@url \@url }%
	\providecommand \@url [1]{\endgroup\@href {#1}{\urlprefix }}%
	\providecommand \urlprefix  [0]{URL }%
	\providecommand \Eprint [0]{\href }%
	\providecommand \doibase [0]{https://doi.org/}%
	\providecommand \selectlanguage [0]{\@gobble}%
	\providecommand \bibinfo  [0]{\@secondoftwo}%
	\providecommand \bibfield  [0]{\@secondoftwo}%
	\providecommand \translation [1]{[#1]}%
	\providecommand \BibitemOpen [0]{}%
	\providecommand \bibitemStop [0]{}%
	\providecommand \bibitemNoStop [0]{.\EOS\space}%
	\providecommand \EOS [0]{\spacefactor3000\relax}%
	\providecommand \BibitemShut  [1]{\csname bibitem#1\endcsname}%
	\let\auto@bib@innerbib\@empty
	\bibitem [{\citenamefont {Balents}(2010)}]{Balents2010}%
	\BibitemOpen
	\bibfield  {author} {\bibinfo {author} {\bibfnamefont {L.}~\bibnamefont
			{Balents}},\ }\bibfield  {title} {\bibinfo {title} {Spin liquids in
			frustrated magnets},\ }\href@noop {} {\bibfield  {journal} {\bibinfo
			{journal} {Nature}\ }\textbf {\bibinfo {volume} {464}},\ \bibinfo {pages}
		{199} (\bibinfo {year} {2010})}\BibitemShut {NoStop}%
	\bibitem [{\citenamefont {Savary}\ and\ \citenamefont
		{Balents}(2016)}]{Savary2016}%
	\BibitemOpen
	\bibfield  {author} {\bibinfo {author} {\bibfnamefont {L.}~\bibnamefont
			{Savary}}\ and\ \bibinfo {author} {\bibfnamefont {L.}~\bibnamefont
			{Balents}},\ }\bibfield  {title} {\bibinfo {title} {Quantum spin liquids: a
			review},\ }\href@noop {} {\bibfield  {journal} {\bibinfo  {journal} {Rep.
				Prog. Phys.}\ }\textbf {\bibinfo {volume} {80}},\ \bibinfo {pages} {016502}
		(\bibinfo {year} {2016})}\BibitemShut {NoStop}%
	\bibitem [{\citenamefont {Anderson}(1973)}]{Anderson1973}%
	\BibitemOpen
	\bibfield  {author} {\bibinfo {author} {\bibfnamefont {P.}~\bibnamefont
			{Anderson}},\ }\bibfield  {title} {\bibinfo {title} {Resonating valence
			bonds: A new kind of insulator?},\ }\href@noop {} {\bibfield  {journal}
		{\bibinfo  {journal} {Mater. Res. Bull.}\ }\textbf {\bibinfo {volume} {8}},\
		\bibinfo {pages} {153} (\bibinfo {year} {1973})}\BibitemShut {NoStop}%
	\bibitem [{\citenamefont {Anderson}(1987)}]{ANDERSON1987}%
	\BibitemOpen
	\bibfield  {author} {\bibinfo {author} {\bibfnamefont {P.~W.}\ \bibnamefont
			{Anderson}},\ }\bibfield  {title} {\bibinfo {title} {The resonating valence
			bond state in La2CuO4 and superconductivity},\ }\href@noop {} {\bibfield
		{journal} {\bibinfo  {journal} {Science}\ }\textbf {\bibinfo {volume}
			{235}},\ \bibinfo {pages} {1196} (\bibinfo {year} {1987})}\BibitemShut
	{NoStop}%
	\bibitem [{\citenamefont {Wen}(1991)}]{Wen1991}%
	\BibitemOpen
	\bibfield  {author} {\bibinfo {author} {\bibfnamefont {X.~G.}\ \bibnamefont
			{Wen}},\ }\bibfield  {title} {\bibinfo {title} {Mean-field theory of
			spin-liquid states with finite energy gap and topological orders},\
	}\href@noop {} {\bibfield  {journal} {\bibinfo  {journal} {Phys. Rev. B}\
		}\textbf {\bibinfo {volume} {44}},\ \bibinfo {pages} {2664} (\bibinfo {year}
		{1991})}\BibitemShut {NoStop}%
	\bibitem [{\citenamefont {Kitaev}(2006)}]{Kitaev2006}%
	\BibitemOpen
	\bibfield  {author} {\bibinfo {author} {\bibfnamefont {A.}~\bibnamefont
			{Kitaev}},\ }\bibfield  {title} {\bibinfo {title} {Anyons in an exactly
			solved model and beyond},\ }\href@noop {} {\bibfield  {journal} {\bibinfo
			{journal} {Ann. Phys.}\ }\textbf {\bibinfo {volume} {321}},\ \bibinfo {pages}
		{2} (\bibinfo {year} {2006})}\BibitemShut {NoStop}%
	\bibitem [{\citenamefont {Shores}\ \emph {et~al.}(2005)\citenamefont {Shores},
		\citenamefont {Nytko}, \citenamefont {Bartlett},\ and\ \citenamefont
		{Nocera}}]{Shores2005}%
	\BibitemOpen
	\bibfield  {author} {\bibinfo {author} {\bibfnamefont {M.~P.}\ \bibnamefont
			{Shores}}, \bibinfo {author} {\bibfnamefont {E.~A.}\ \bibnamefont {Nytko}},
		\bibinfo {author} {\bibfnamefont {B.~M.}\ \bibnamefont {Bartlett}},\ and\
		\bibinfo {author} {\bibfnamefont {D.~G.}\ \bibnamefont {Nocera}},\ }\bibfield
	{title} {\bibinfo {title} {A structurally perfect s=1/2 kagom{\'{e}}
			antiferromagnet},\ }\href@noop {} {\bibfield  {journal} {\bibinfo  {journal}
			{J. Am. Chem. Soc}\ }\textbf {\bibinfo {volume} {127}},\ \bibinfo {pages}
		{13462} (\bibinfo {year} {2005})}\BibitemShut {NoStop}%
	\bibitem [{\citenamefont {Han}\ \emph {et~al.}(2012)\citenamefont {Han},
		\citenamefont {Helton}, \citenamefont {Chu}, \citenamefont {Nocera},
		\citenamefont {Rodriguez-Rivera}, \citenamefont {Broholm},\ and\
		\citenamefont {Lee}}]{Han2012}%
	\BibitemOpen
	\bibfield  {author} {\bibinfo {author} {\bibfnamefont {T.-H.}\ \bibnamefont
			{Han}}, \bibinfo {author} {\bibfnamefont {J.~S.}\ \bibnamefont {Helton}},
		\bibinfo {author} {\bibfnamefont {S.}~\bibnamefont {Chu}}, \bibinfo {author}
		{\bibfnamefont {D.~G.}\ \bibnamefont {Nocera}}, \bibinfo {author}
		{\bibfnamefont {J.~A.}\ \bibnamefont {Rodriguez-Rivera}}, \bibinfo {author}
		{\bibfnamefont {C.}~\bibnamefont {Broholm}},\ and\ \bibinfo {author}
		{\bibfnamefont {Y.~S.}\ \bibnamefont {Lee}},\ }\bibfield  {title} {\bibinfo
		{title} {Fractionalized excitations in the spin-liquid state of a
			kagome-lattice antiferromagnet},\ }\href@noop {} {\bibfield  {journal}
		{\bibinfo  {journal} {Nature}\ }\textbf {\bibinfo {volume} {492}},\ \bibinfo
		{pages} {406} (\bibinfo {year} {2012})}\BibitemShut {NoStop}%
	\bibitem [{\citenamefont {Li}\ \emph {et~al.}(2015{\natexlab{a}})\citenamefont
		{Li}, \citenamefont {Chen}, \citenamefont {Tong}, \citenamefont {Pi},
		\citenamefont {Liu}, \citenamefont {Yang}, \citenamefont {Wang},\ and\
		\citenamefont {Zhang}}]{Li2015a}%
	\BibitemOpen
	\bibfield  {author} {\bibinfo {author} {\bibfnamefont {Y.}~\bibnamefont
			{Li}}, \bibinfo {author} {\bibfnamefont {G.}~\bibnamefont {Chen}}, \bibinfo
		{author} {\bibfnamefont {W.}~\bibnamefont {Tong}}, \bibinfo {author}
		{\bibfnamefont {L.}~\bibnamefont {Pi}}, \bibinfo {author} {\bibfnamefont
			{J.}~\bibnamefont {Liu}}, \bibinfo {author} {\bibfnamefont {Z.}~\bibnamefont
			{Yang}}, \bibinfo {author} {\bibfnamefont {X.}~\bibnamefont {Wang}},\ and\
		\bibinfo {author} {\bibfnamefont {Q.}~\bibnamefont {Zhang}},\ }\bibfield
	{title} {\bibinfo {title} {Rare-earth triangular lattice spin liquid: A
			single-crystal study of \ce{YbMgGaO4}},\ }\href@noop {} {\bibfield  {journal}
		{\bibinfo  {journal} {Phys. Rev. Lett.}\ }\textbf {\bibinfo {volume} {115}}
		(\bibinfo {year} {2015}{\natexlab{a}})}\BibitemShut {NoStop}%
	\bibitem [{\citenamefont {Li}\ \emph {et~al.}(2015{\natexlab{b}})\citenamefont
		{Li}, \citenamefont {Liao}, \citenamefont {Zhang}, \citenamefont {Li},
		\citenamefont {Jin}, \citenamefont {Ling}, \citenamefont {Zhang},
		\citenamefont {Zou}, \citenamefont {Pi}, \citenamefont {Yang}, \citenamefont
		{Wang}, \citenamefont {Wu},\ and\ \citenamefont {Zhang}}]{Li2015}%
	\BibitemOpen
	\bibfield  {author} {\bibinfo {author} {\bibfnamefont {Y.}~\bibnamefont
			{Li}}, \bibinfo {author} {\bibfnamefont {H.}~\bibnamefont {Liao}}, \bibinfo
		{author} {\bibfnamefont {Z.}~\bibnamefont {Zhang}}, \bibinfo {author}
		{\bibfnamefont {S.}~\bibnamefont {Li}}, \bibinfo {author} {\bibfnamefont
			{F.}~\bibnamefont {Jin}}, \bibinfo {author} {\bibfnamefont {L.}~\bibnamefont
			{Ling}}, \bibinfo {author} {\bibfnamefont {L.}~\bibnamefont {Zhang}},
		\bibinfo {author} {\bibfnamefont {Y.}~\bibnamefont {Zou}}, \bibinfo {author}
		{\bibfnamefont {L.}~\bibnamefont {Pi}}, \bibinfo {author} {\bibfnamefont
			{Z.}~\bibnamefont {Yang}}, \bibinfo {author} {\bibfnamefont {J.}~\bibnamefont
			{Wang}}, \bibinfo {author} {\bibfnamefont {Z.}~\bibnamefont {Wu}},\ and\
		\bibinfo {author} {\bibfnamefont {Q.}~\bibnamefont {Zhang}},\ }\bibfield
	{title} {\bibinfo {title} {Gapless quantum spin liquid ground state in the
			two-dimensional spin-1/2 triangular antiferromagnet \ce{YbMgGaO4}},\
	}\href@noop {} {\bibfield  {journal} {\bibinfo  {journal} {Sci. Rep.}\
		}\textbf {\bibinfo {volume} {5}} (\bibinfo {year}
		{2015}{\natexlab{b}})}\BibitemShut {NoStop}%
	\bibitem [{\citenamefont {Li}\ \emph {et~al.}(2016)\citenamefont {Li},
		\citenamefont {Adroja}, \citenamefont {Biswas}, \citenamefont {Baker},
		\citenamefont {Zhang}, \citenamefont {Liu}, \citenamefont {Tsirlin},
		\citenamefont {Gegenwart},\ and\ \citenamefont {Zhang}}]{Li2016a}%
	\BibitemOpen
	\bibfield  {author} {\bibinfo {author} {\bibfnamefont {Y.}~\bibnamefont
			{Li}}, \bibinfo {author} {\bibfnamefont {D.}~\bibnamefont {Adroja}}, \bibinfo
		{author} {\bibfnamefont {P.~K.}\ \bibnamefont {Biswas}}, \bibinfo {author}
		{\bibfnamefont {P.~J.}\ \bibnamefont {Baker}}, \bibinfo {author}
		{\bibfnamefont {Q.}~\bibnamefont {Zhang}}, \bibinfo {author} {\bibfnamefont
			{J.}~\bibnamefont {Liu}}, \bibinfo {author} {\bibfnamefont {A.~A.}\
			\bibnamefont {Tsirlin}}, \bibinfo {author} {\bibfnamefont {P.}~\bibnamefont
			{Gegenwart}},\ and\ \bibinfo {author} {\bibfnamefont {Q.}~\bibnamefont
			{Zhang}},\ }\bibfield  {title} {\bibinfo {title} {Muon spin relaxation
			evidence for the u(1) quantum spin-liquid ground state in the triangular
			antiferromagnet \ce{YbMgGaO4}},\ }\href@noop {} {\bibfield  {journal}
		{\bibinfo  {journal} {Phys. Rev. Lett.}\ }\textbf {\bibinfo {volume} {117}}
		(\bibinfo {year} {2016})}\BibitemShut {NoStop}%
	\bibitem [{\citenamefont {Li}\ \emph {et~al.}(2017{\natexlab{a}})\citenamefont
		{Li}, \citenamefont {Adroja}, \citenamefont {Voneshen}, \citenamefont
		{Bewley}, \citenamefont {Zhang}, \citenamefont {Tsirlin},\ and\ \citenamefont
		{Gegenwart}}]{Li2017}%
	\BibitemOpen
	\bibfield  {author} {\bibinfo {author} {\bibfnamefont {Y.}~\bibnamefont
			{Li}}, \bibinfo {author} {\bibfnamefont {D.}~\bibnamefont {Adroja}}, \bibinfo
		{author} {\bibfnamefont {D.}~\bibnamefont {Voneshen}}, \bibinfo {author}
		{\bibfnamefont {R.~I.}\ \bibnamefont {Bewley}}, \bibinfo {author}
		{\bibfnamefont {Q.}~\bibnamefont {Zhang}}, \bibinfo {author} {\bibfnamefont
			{A.~A.}\ \bibnamefont {Tsirlin}},\ and\ \bibinfo {author} {\bibfnamefont
			{P.}~\bibnamefont {Gegenwart}},\ }\bibfield  {title} {\bibinfo {title}
		{Nearest-neighbour resonating valence bonds in \ce{YbMgGaO4}},\ }\href@noop
	{} {\bibfield  {journal} {\bibinfo  {journal} {Nat. Commun.}\ }\textbf
		{\bibinfo {volume} {8}} (\bibinfo {year} {2017}{\natexlab{a}})}\BibitemShut
	{NoStop}%
	\bibitem [{\citenamefont {Shen}\ \emph {et~al.}(2016)\citenamefont {Shen},
		\citenamefont {Li}, \citenamefont {Wo}, \citenamefont {Li}, \citenamefont
		{Shen}, \citenamefont {Pan}, \citenamefont {Wang}, \citenamefont {Walker},
		\citenamefont {Steffens}, \citenamefont {Boehm}, \citenamefont {Hao},
		\citenamefont {Quintero-Castro}, \citenamefont {Harriger}, \citenamefont
		{Frontzek}, \citenamefont {Hao}, \citenamefont {Meng}, \citenamefont {Zhang},
		\citenamefont {Chen},\ and\ \citenamefont {Zhao}}]{Shen2016}%
	\BibitemOpen
	\bibfield  {author} {\bibinfo {author} {\bibfnamefont {Y.}~\bibnamefont
			{Shen}}, \bibinfo {author} {\bibfnamefont {Y.-D.}\ \bibnamefont {Li}},
		\bibinfo {author} {\bibfnamefont {H.}~\bibnamefont {Wo}}, \bibinfo {author}
		{\bibfnamefont {Y.}~\bibnamefont {Li}}, \bibinfo {author} {\bibfnamefont
			{S.}~\bibnamefont {Shen}}, \bibinfo {author} {\bibfnamefont {B.}~\bibnamefont
			{Pan}}, \bibinfo {author} {\bibfnamefont {Q.}~\bibnamefont {Wang}}, \bibinfo
		{author} {\bibfnamefont {H.~C.}\ \bibnamefont {Walker}}, \bibinfo {author}
		{\bibfnamefont {P.}~\bibnamefont {Steffens}}, \bibinfo {author}
		{\bibfnamefont {M.}~\bibnamefont {Boehm}}, \bibinfo {author} {\bibfnamefont
			{Y.}~\bibnamefont {Hao}}, \bibinfo {author} {\bibfnamefont {D.~L.}\
			\bibnamefont {Quintero-Castro}}, \bibinfo {author} {\bibfnamefont {L.~W.}\
			\bibnamefont {Harriger}}, \bibinfo {author} {\bibfnamefont {M.~D.}\
			\bibnamefont {Frontzek}}, \bibinfo {author} {\bibfnamefont {L.}~\bibnamefont
			{Hao}}, \bibinfo {author} {\bibfnamefont {S.}~\bibnamefont {Meng}}, \bibinfo
		{author} {\bibfnamefont {Q.}~\bibnamefont {Zhang}}, \bibinfo {author}
		{\bibfnamefont {G.}~\bibnamefont {Chen}},\ and\ \bibinfo {author}
		{\bibfnamefont {J.}~\bibnamefont {Zhao}},\ }\bibfield  {title} {\bibinfo
		{title} {Evidence for a spinon fermi surface in a triangular-lattice
			quantum-spin-liquid candidate},\ }\href@noop {} {\bibfield  {journal}
		{\bibinfo  {journal} {Nature}\ }\textbf {\bibinfo {volume} {540}},\ \bibinfo
		{pages} {559} (\bibinfo {year} {2016})}\BibitemShut {NoStop}%
	\bibitem [{\citenamefont {Li}(2019)}]{Li2019}%
	\BibitemOpen
	\bibfield  {author} {\bibinfo {author} {\bibfnamefont {Y.}~\bibnamefont
			{Li}},\ }\bibfield  {title} {\bibinfo {title} {\ce{YbMgGaO4} : A
			triangular-lattice quantum spin liquid candidate},\ }\href@noop {} {\bibfield
		{journal} {\bibinfo  {journal} {Adv. Quantum Technol}\ }\textbf {\bibinfo
			{volume} {2}},\ \bibinfo {pages} {1900089} (\bibinfo {year}
		{2019})}\BibitemShut {NoStop}%
	\bibitem [{\citenamefont {Li}\ \emph {et~al.}(2017{\natexlab{b}})\citenamefont
		{Li}, \citenamefont {Adroja}, \citenamefont {Bewley}, \citenamefont
		{Voneshen}, \citenamefont {Tsirlin}, \citenamefont {Gegenwart},\ and\
		\citenamefont {Zhang}}]{LiYuSheng_CEF}%
	\BibitemOpen
	\bibfield  {author} {\bibinfo {author} {\bibfnamefont {Y.}~\bibnamefont
			{Li}}, \bibinfo {author} {\bibfnamefont {D.}~\bibnamefont {Adroja}}, \bibinfo
		{author} {\bibfnamefont {R.~I.}\ \bibnamefont {Bewley}}, \bibinfo {author}
		{\bibfnamefont {D.}~\bibnamefont {Voneshen}}, \bibinfo {author}
		{\bibfnamefont {A.~A.}\ \bibnamefont {Tsirlin}}, \bibinfo {author}
		{\bibfnamefont {P.}~\bibnamefont {Gegenwart}},\ and\ \bibinfo {author}
		{\bibfnamefont {Q.}~\bibnamefont {Zhang}},\ }\href@noop {} {\bibfield
		{journal} {\bibinfo  {journal} {Phys. Rev. Lett.}\ }\textbf {\bibinfo
			{volume} {118}},\ \bibinfo {pages} {107202} (\bibinfo {year}
		{2017}{\natexlab{b}})}\BibitemShut {NoStop}%
	\bibitem [{\citenamefont {Liu}\ \emph {et~al.}(2018)\citenamefont {Liu},
		\citenamefont {Zhang}, \citenamefont {Ji}, \citenamefont {Liu}, \citenamefont
		{Li}, \citenamefont {Wang}, \citenamefont {Lei}, \citenamefont {Chen},\ and\
		\citenamefont {Zhang}}]{Liu2018}%
	\BibitemOpen
	\bibfield  {author} {\bibinfo {author} {\bibfnamefont {W.}~\bibnamefont
			{Liu}}, \bibinfo {author} {\bibfnamefont {Z.}~\bibnamefont {Zhang}}, \bibinfo
		{author} {\bibfnamefont {J.}~\bibnamefont {Ji}}, \bibinfo {author}
		{\bibfnamefont {Y.}~\bibnamefont {Liu}}, \bibinfo {author} {\bibfnamefont
			{J.}~\bibnamefont {Li}}, \bibinfo {author} {\bibfnamefont {X.}~\bibnamefont
			{Wang}}, \bibinfo {author} {\bibfnamefont {H.}~\bibnamefont {Lei}}, \bibinfo
		{author} {\bibfnamefont {G.}~\bibnamefont {Chen}},\ and\ \bibinfo {author}
		{\bibfnamefont {Q.}~\bibnamefont {Zhang}},\ }\bibfield  {title} {\bibinfo
		{title} {Rare-earth chalcogenides: A large family of triangular lattice spin
			liquid candidates},\ }\href@noop {} {\bibfield  {journal} {\bibinfo
			{journal} {Chin. Phys. Lett.}\ }\textbf {\bibinfo {volume} {35}},\ \bibinfo
		{pages} {117501} (\bibinfo {year} {2018})}\BibitemShut {NoStop}%
	\bibitem [{\citenamefont {Baenitz}\ \emph {et~al.}(2018)\citenamefont
		{Baenitz}, \citenamefont {Schlender}, \citenamefont {Sichelschmidt},
		\citenamefont {Onykiienko}, \citenamefont {Zangeneh}, \citenamefont
		{Ranjith}, \citenamefont {Sarkar}, \citenamefont {Hozoi}, \citenamefont
		{Walker}, \citenamefont {Orain}, \citenamefont {Yasuoka}, \citenamefont
		{van~den Brink}, \citenamefont {Klauss}, \citenamefont {Inosov},\ and\
		\citenamefont {Doert}}]{Baenitz2018}%
	\BibitemOpen
	\bibfield  {author} {\bibinfo {author} {\bibfnamefont {M.}~\bibnamefont
			{Baenitz}}, \bibinfo {author} {\bibfnamefont {P.}~\bibnamefont {Schlender}},
		\bibinfo {author} {\bibfnamefont {J.}~\bibnamefont {Sichelschmidt}}, \bibinfo
		{author} {\bibfnamefont {Y.~A.}\ \bibnamefont {Onykiienko}}, \bibinfo
		{author} {\bibfnamefont {Z.}~\bibnamefont {Zangeneh}}, \bibinfo {author}
		{\bibfnamefont {K.~M.}\ \bibnamefont {Ranjith}}, \bibinfo {author}
		{\bibfnamefont {R.}~\bibnamefont {Sarkar}}, \bibinfo {author} {\bibfnamefont
			{L.}~\bibnamefont {Hozoi}}, \bibinfo {author} {\bibfnamefont {H.~C.}\
			\bibnamefont {Walker}}, \bibinfo {author} {\bibfnamefont {J.-C.}\
			\bibnamefont {Orain}}, \bibinfo {author} {\bibfnamefont {H.}~\bibnamefont
			{Yasuoka}}, \bibinfo {author} {\bibfnamefont {J.}~\bibnamefont {van~den
				Brink}}, \bibinfo {author} {\bibfnamefont {H.~H.}\ \bibnamefont {Klauss}},
		\bibinfo {author} {\bibfnamefont {D.~S.}\ \bibnamefont {Inosov}},\ and\
		\bibinfo {author} {\bibfnamefont {T.}~\bibnamefont {Doert}},\ }\bibfield
	{title} {\bibinfo {title} {\ce{NaYbS2}: A planar spin-1/2 triangular-lattice
			magnet and putative spin liquid},\ }\href@noop {} {\bibfield  {journal}
		{\bibinfo  {journal} {Phys. Rev. B}\ }\textbf {\bibinfo {volume} {98}},\
		\bibinfo {pages} {220409} (\bibinfo {year} {2018})}\BibitemShut {NoStop}%
	\bibitem [{\citenamefont {Ranjith}\ \emph
		{et~al.}(2019{\natexlab{a}})\citenamefont {Ranjith}, \citenamefont
		{Dmytriieva}, \citenamefont {Khim}, \citenamefont {Sichelschmidt},
		\citenamefont {Luther}, \citenamefont {Ehlers}, \citenamefont {Yasuoka},
		\citenamefont {Wosnitza}, \citenamefont {Tsirlin}, \citenamefont {Kühne},\
		and\ \citenamefont {Baenitz}}]{Ranjith2019}%
	\BibitemOpen
	\bibfield  {author} {\bibinfo {author} {\bibfnamefont {K.~M.}\ \bibnamefont
			{Ranjith}}, \bibinfo {author} {\bibfnamefont {D.}~\bibnamefont {Dmytriieva}},
		\bibinfo {author} {\bibfnamefont {S.}~\bibnamefont {Khim}}, \bibinfo {author}
		{\bibfnamefont {J.}~\bibnamefont {Sichelschmidt}}, \bibinfo {author}
		{\bibfnamefont {S.}~\bibnamefont {Luther}}, \bibinfo {author} {\bibfnamefont
			{D.}~\bibnamefont {Ehlers}}, \bibinfo {author} {\bibfnamefont
			{H.}~\bibnamefont {Yasuoka}}, \bibinfo {author} {\bibfnamefont
			{J.}~\bibnamefont {Wosnitza}}, \bibinfo {author} {\bibfnamefont {A.~A.}\
			\bibnamefont {Tsirlin}}, \bibinfo {author} {\bibfnamefont {H.}~\bibnamefont
			{Kühne}},\ and\ \bibinfo {author} {\bibfnamefont {M.}~\bibnamefont
			{Baenitz}},\ }\bibfield  {title} {\bibinfo {title} {Field-induced instability
			of the quantum spin liquid ground state in the jeff=12 triangular-lattice
			compound \ce{NaYbO2}},\ }\href@noop {} {\bibfield  {journal} {\bibinfo
			{journal} {Phys. Rev. B}\ }\textbf {\bibinfo {volume} {99}} (\bibinfo {year}
		{2019}{\natexlab{a}})}\BibitemShut {NoStop}%
	\bibitem [{\citenamefont {Ding}\ \emph {et~al.}(2019)\citenamefont {Ding},
		\citenamefont {Manuel}, \citenamefont {Bachus}, \citenamefont {Gru{\ss}ler},
		\citenamefont {Gegenwart}, \citenamefont {Singleton}, \citenamefont
		{Johnson}, \citenamefont {Walker}, \citenamefont {Adroja}, \citenamefont
		{Hillier},\ and\ \citenamefont {Tsirlin}}]{Ding2019a}%
	\BibitemOpen
	\bibfield  {author} {\bibinfo {author} {\bibfnamefont {L.}~\bibnamefont
			{Ding}}, \bibinfo {author} {\bibfnamefont {P.}~\bibnamefont {Manuel}},
		\bibinfo {author} {\bibfnamefont {S.}~\bibnamefont {Bachus}}, \bibinfo
		{author} {\bibfnamefont {F.}~\bibnamefont {Gru{\ss}ler}}, \bibinfo {author}
		{\bibfnamefont {P.}~\bibnamefont {Gegenwart}}, \bibinfo {author}
		{\bibfnamefont {J.}~\bibnamefont {Singleton}}, \bibinfo {author}
		{\bibfnamefont {R.~D.}\ \bibnamefont {Johnson}}, \bibinfo {author}
		{\bibfnamefont {H.~C.}\ \bibnamefont {Walker}}, \bibinfo {author}
		{\bibfnamefont {D.~T.}\ \bibnamefont {Adroja}}, \bibinfo {author}
		{\bibfnamefont {A.~D.}\ \bibnamefont {Hillier}},\ and\ \bibinfo {author}
		{\bibfnamefont {A.~A.}\ \bibnamefont {Tsirlin}},\ }\bibfield  {title}
	{\bibinfo {title} {Gapless spin-liquid state in the structurally
			disorder-free triangular antiferromagnet \ce{NaYbO2}},\ }\href@noop {}
	{\bibfield  {journal} {\bibinfo  {journal} {Phys. Rev. B}\ }\textbf {\bibinfo
			{volume} {100}} (\bibinfo {year} {2019})}\BibitemShut {NoStop}%
	\bibitem [{\citenamefont {Sichelschmidt}\ \emph {et~al.}(2019)\citenamefont
		{Sichelschmidt}, \citenamefont {Schlender}, \citenamefont {Schmidt},
		\citenamefont {Baenitz},\ and\ \citenamefont {Doert}}]{Sichelschmidt2019}%
	\BibitemOpen
	\bibfield  {author} {\bibinfo {author} {\bibfnamefont {J.}~\bibnamefont
			{Sichelschmidt}}, \bibinfo {author} {\bibfnamefont {P.}~\bibnamefont
			{Schlender}}, \bibinfo {author} {\bibfnamefont {B.}~\bibnamefont {Schmidt}},
		\bibinfo {author} {\bibfnamefont {M.}~\bibnamefont {Baenitz}},\ and\ \bibinfo
		{author} {\bibfnamefont {T.}~\bibnamefont {Doert}},\ }\bibfield  {title}
	{\bibinfo {title} {Electron spin resonance on the spin-1/2 triangular magnet
			\ce{NaYbS2}},\ }\href@noop {} {\bibfield  {journal} {\bibinfo  {journal} {J.
				Phys.: Condens. Matter}\ }\textbf {\bibinfo {volume} {31}},\ \bibinfo {pages}
		{205601} (\bibinfo {year} {2019})}\BibitemShut {NoStop}%
	\bibitem [{\citenamefont {Ranjith}\ \emph
		{et~al.}(2019{\natexlab{b}})\citenamefont {Ranjith}, \citenamefont {Luther},
		\citenamefont {Reimann}, \citenamefont {Schmidt}, \citenamefont {Schlender},
		\citenamefont {Sichelschmidt}, \citenamefont {Yasuoka}, \citenamefont
		{Strydom}, \citenamefont {Skourski}, \citenamefont {Wosnitza}, \citenamefont
		{Kühne}, \citenamefont {Doert},\ and\ \citenamefont
		{Baenitz}}]{Ranjith2019a}%
	\BibitemOpen
	\bibfield  {author} {\bibinfo {author} {\bibfnamefont {K.~M.}\ \bibnamefont
			{Ranjith}}, \bibinfo {author} {\bibfnamefont {S.}~\bibnamefont {Luther}},
		\bibinfo {author} {\bibfnamefont {T.}~\bibnamefont {Reimann}}, \bibinfo
		{author} {\bibfnamefont {B.}~\bibnamefont {Schmidt}}, \bibinfo {author}
		{\bibfnamefont {P.}~\bibnamefont {Schlender}}, \bibinfo {author}
		{\bibfnamefont {J.}~\bibnamefont {Sichelschmidt}}, \bibinfo {author}
		{\bibfnamefont {H.}~\bibnamefont {Yasuoka}}, \bibinfo {author} {\bibfnamefont
			{A.~M.}\ \bibnamefont {Strydom}}, \bibinfo {author} {\bibfnamefont
			{Y.}~\bibnamefont {Skourski}}, \bibinfo {author} {\bibfnamefont
			{J.}~\bibnamefont {Wosnitza}}, \bibinfo {author} {\bibfnamefont
			{H.}~\bibnamefont {Kühne}}, \bibinfo {author} {\bibfnamefont
			{T.}~\bibnamefont {Doert}},\ and\ \bibinfo {author} {\bibfnamefont
			{M.}~\bibnamefont {Baenitz}},\ }\bibfield  {title} {\bibinfo {title}
		{Anisotropic field-induced ordering in the triangular-lattice quantum spin
			liquid \ce{NaYbSe2}},\ }\href@noop {} {\bibfield  {journal} {\bibinfo
			{journal} {Phys. Rev. B}\ }\textbf {\bibinfo {volume} {100}} (\bibinfo {year}
		{2019}{\natexlab{b}})}\BibitemShut {NoStop}%
	\bibitem [{\citenamefont {Xing}\ \emph
		{et~al.}(2019{\natexlab{a}})\citenamefont {Xing}, \citenamefont {Sanjeewa},
		\citenamefont {Kim}, \citenamefont {Stewart}, \citenamefont {Du},
		\citenamefont {Reboredo}, \citenamefont {Custelcean},\ and\ \citenamefont
		{Sefat}}]{Xing2019}%
	\BibitemOpen
	\bibfield  {author} {\bibinfo {author} {\bibfnamefont {J.}~\bibnamefont
			{Xing}}, \bibinfo {author} {\bibfnamefont {L.~D.}\ \bibnamefont {Sanjeewa}},
		\bibinfo {author} {\bibfnamefont {J.}~\bibnamefont {Kim}}, \bibinfo {author}
		{\bibfnamefont {G.~R.}\ \bibnamefont {Stewart}}, \bibinfo {author}
		{\bibfnamefont {M.-H.}\ \bibnamefont {Du}}, \bibinfo {author} {\bibfnamefont
			{F.~A.}\ \bibnamefont {Reboredo}}, \bibinfo {author} {\bibfnamefont
			{R.}~\bibnamefont {Custelcean}},\ and\ \bibinfo {author} {\bibfnamefont
			{A.~S.}\ \bibnamefont {Sefat}},\ }\bibfield  {title} {\bibinfo {title}
		{Crystal synthesis and frustrated magnetism in triangular lattice
			\ce{CsRESe2} (\ce{RE} = la{\textendash}lu): Quantum spin liquid candidates
			\ce{CsCeSe2} and \ce{CsYbSe2}},\ }\href@noop {} {\bibfield  {journal}
		{\bibinfo  {journal} {Mater. Lett}\ ,\ \bibinfo {pages} {71}} (\bibinfo
		{year} {2019}{\natexlab{a}})}\BibitemShut {NoStop}%
	\bibitem [{\citenamefont {Bordelon}\ \emph {et~al.}(2019)\citenamefont
		{Bordelon}, \citenamefont {Kenney}, \citenamefont {Liu}, \citenamefont
		{Hogan}, \citenamefont {Posthuma}, \citenamefont {Kavand}, \citenamefont
		{Lyu}, \citenamefont {Sherwin}, \citenamefont {Butch}, \citenamefont {Brown},
		\citenamefont {Graf}, \citenamefont {Balents},\ and\ \citenamefont
		{Wilson}}]{Bordelon2019}%
	\BibitemOpen
	\bibfield  {author} {\bibinfo {author} {\bibfnamefont {M.~M.}\ \bibnamefont
			{Bordelon}}, \bibinfo {author} {\bibfnamefont {E.}~\bibnamefont {Kenney}},
		\bibinfo {author} {\bibfnamefont {C.}~\bibnamefont {Liu}}, \bibinfo {author}
		{\bibfnamefont {T.}~\bibnamefont {Hogan}}, \bibinfo {author} {\bibfnamefont
			{L.}~\bibnamefont {Posthuma}}, \bibinfo {author} {\bibfnamefont
			{M.}~\bibnamefont {Kavand}}, \bibinfo {author} {\bibfnamefont
			{Y.}~\bibnamefont {Lyu}}, \bibinfo {author} {\bibfnamefont {M.}~\bibnamefont
			{Sherwin}}, \bibinfo {author} {\bibfnamefont {N.~P.}\ \bibnamefont {Butch}},
		\bibinfo {author} {\bibfnamefont {C.}~\bibnamefont {Brown}}, \bibinfo
		{author} {\bibfnamefont {M.~J.}\ \bibnamefont {Graf}}, \bibinfo {author}
		{\bibfnamefont {L.}~\bibnamefont {Balents}},\ and\ \bibinfo {author}
		{\bibfnamefont {S.~D.}\ \bibnamefont {Wilson}},\ }\bibfield  {title}
	{\bibinfo {title} {Field-tunable quantum disordered ground state in the
			triangular-lattice antiferromagnet \ce{NaYbO2}},\ }\href@noop {} {\bibfield
		{journal} {\bibinfo  {journal} {Nat. Phys}\ }\textbf {\bibinfo {volume}
			{15}},\ \bibinfo {pages} {1058} (\bibinfo {year} {2019})}\BibitemShut
	{NoStop}%
	\bibitem [{\citenamefont {Xing}\ \emph
		{et~al.}(2019{\natexlab{b}})\citenamefont {Xing}, \citenamefont {Sanjeewa},
		\citenamefont {Kim}, \citenamefont {Meier}, \citenamefont {May},
		\citenamefont {Zheng}, \citenamefont {Custelcean}, \citenamefont {Stewart},\
		and\ \citenamefont {Sefat}}]{Xing2019a}%
	\BibitemOpen
	\bibfield  {author} {\bibinfo {author} {\bibfnamefont {J.}~\bibnamefont
			{Xing}}, \bibinfo {author} {\bibfnamefont {L.~D.}\ \bibnamefont {Sanjeewa}},
		\bibinfo {author} {\bibfnamefont {J.}~\bibnamefont {Kim}}, \bibinfo {author}
		{\bibfnamefont {W.~R.}\ \bibnamefont {Meier}}, \bibinfo {author}
		{\bibfnamefont {A.~F.}\ \bibnamefont {May}}, \bibinfo {author} {\bibfnamefont
			{Q.}~\bibnamefont {Zheng}}, \bibinfo {author} {\bibfnamefont
			{R.}~\bibnamefont {Custelcean}}, \bibinfo {author} {\bibfnamefont {G.~R.}\
			\bibnamefont {Stewart}},\ and\ \bibinfo {author} {\bibfnamefont {A.~S.}\
			\bibnamefont {Sefat}},\ }\bibfield  {title} {\bibinfo {title} {Synthesis,
			magnetization, and heat capacity of triangular lattice materials {NaErSe}2
			and {KErSe}2},\ }\href@noop {} {\bibfield  {journal} {\bibinfo  {journal}
			{Phys. Rev. Mater}\ }\textbf {\bibinfo {volume} {3}} (\bibinfo {year}
		{2019}{\natexlab{b}})}\BibitemShut {NoStop}%
	\bibitem [{\citenamefont {Xing}\ \emph
		{et~al.}(2019{\natexlab{c}})\citenamefont {Xing}, \citenamefont {Sanjeewa},
		\citenamefont {Kim}, \citenamefont {Stewart}, \citenamefont {Podlesnyak},\
		and\ \citenamefont {Sefat}}]{Xing2019b}%
	\BibitemOpen
	\bibfield  {author} {\bibinfo {author} {\bibfnamefont {J.}~\bibnamefont
			{Xing}}, \bibinfo {author} {\bibfnamefont {L.~D.}\ \bibnamefont {Sanjeewa}},
		\bibinfo {author} {\bibfnamefont {J.}~\bibnamefont {Kim}}, \bibinfo {author}
		{\bibfnamefont {G.~R.}\ \bibnamefont {Stewart}}, \bibinfo {author}
		{\bibfnamefont {A.}~\bibnamefont {Podlesnyak}},\ and\ \bibinfo {author}
		{\bibfnamefont {A.~S.}\ \bibnamefont {Sefat}},\ }\bibfield  {title} {\bibinfo
		{title} {Field-induced magnetic transition and spin fluctuations in the
			quantum spin-liquid candidate \ce{CsYbSe2}},\ }\href@noop {} {\bibfield
		{journal} {\bibinfo  {journal} {Phys. Rev. B}\ }\textbf {\bibinfo {volume}
			{100}} (\bibinfo {year} {2019}{\natexlab{c}})}\BibitemShut {NoStop}%
	\bibitem [{\citenamefont {Gao}\ \emph {et~al.}()\citenamefont {Gao},
		\citenamefont {Xiao}, \citenamefont {Kamazawa}, \citenamefont {Ikeuchi},
		\citenamefont {Biner}, \citenamefont {Krämer}, \citenamefont {Rüegg},\ and\
		\citenamefont {hisa Arima}}]{Gao2019}%
	\BibitemOpen
	\bibfield  {author} {\bibinfo {author} {\bibfnamefont {S.}~\bibnamefont
			{Gao}}, \bibinfo {author} {\bibfnamefont {F.}~\bibnamefont {Xiao}}, \bibinfo
		{author} {\bibfnamefont {K.}~\bibnamefont {Kamazawa}}, \bibinfo {author}
		{\bibfnamefont {K.}~\bibnamefont {Ikeuchi}}, \bibinfo {author} {\bibfnamefont
			{D.}~\bibnamefont {Biner}}, \bibinfo {author} {\bibfnamefont
			{K.}~\bibnamefont {Krämer}}, \bibinfo {author} {\bibfnamefont
			{C.}~\bibnamefont {Rüegg}},\ and\ \bibinfo {author} {\bibfnamefont
			{T.}~\bibnamefont {hisa Arima}},\ }\bibfield  {title} {\bibinfo {title}
		{Crystal-electric-field excitations in a quantum-spin-liquid candidate
			\ce{NaErS2}},\ }\href@noop {} {\ }\Eprint
	{https://arxiv.org/abs/http://arxiv.org/abs/1911.10662v1}
	{http://arxiv.org/abs/1911.10662v1} \BibitemShut {NoStop}%
	\bibitem [{\citenamefont {Sichelschmidt}\ \emph {et~al.}()\citenamefont
		{Sichelschmidt}, \citenamefont {Schmidt}, \citenamefont {Schlender},
		\citenamefont {Khim}, \citenamefont {Doert},\ and\ \citenamefont
		{Baenitz}}]{Sichelschmidt2019a}%
	\BibitemOpen
	\bibfield  {author} {\bibinfo {author} {\bibfnamefont {J.}~\bibnamefont
			{Sichelschmidt}}, \bibinfo {author} {\bibfnamefont {B.}~\bibnamefont
			{Schmidt}}, \bibinfo {author} {\bibfnamefont {P.}~\bibnamefont {Schlender}},
		\bibinfo {author} {\bibfnamefont {S.}~\bibnamefont {Khim}}, \bibinfo {author}
		{\bibfnamefont {T.}~\bibnamefont {Doert}},\ and\ \bibinfo {author}
		{\bibfnamefont {M.}~\bibnamefont {Baenitz}},\ }\bibfield  {title} {\bibinfo
		{title} {Effective spin-1/2 moments on a \ce{Yb^3+} triangular lattice: an
			esr study},\ }\href@noop {} {\ }\Eprint
	{https://arxiv.org/abs/http://arxiv.org/abs/1912.01868v1}
	{http://arxiv.org/abs/1912.01868v1} \BibitemShut {NoStop}%
	\bibitem [{\citenamefont {Sarkar}\ \emph {et~al.}()\citenamefont {Sarkar},
		\citenamefont {Schlender}, \citenamefont {Grinenko}, \citenamefont
		{Haeussler}, \citenamefont {Baker}, \citenamefont {Doert},\ and\
		\citenamefont {Klauss}}]{Sarkar2019}%
	\BibitemOpen
	\bibfield  {author} {\bibinfo {author} {\bibfnamefont {R.}~\bibnamefont
			{Sarkar}}, \bibinfo {author} {\bibfnamefont {P.}~\bibnamefont {Schlender}},
		\bibinfo {author} {\bibfnamefont {V.}~\bibnamefont {Grinenko}}, \bibinfo
		{author} {\bibfnamefont {E.}~\bibnamefont {Haeussler}}, \bibinfo {author}
		{\bibfnamefont {P.~J.}\ \bibnamefont {Baker}}, \bibinfo {author}
		{\bibfnamefont {T.}~\bibnamefont {Doert}},\ and\ \bibinfo {author}
		{\bibfnamefont {H.~H.}\ \bibnamefont {Klauss}},\ }\bibfield  {title}
	{\bibinfo {title} {Quantum spin liquid ground state in the disorder free
			triangular lattice \ce{NaYbS2}},\ }\href@noop {} {\ }\Eprint
	{https://arxiv.org/abs/http://arxiv.org/abs/1911.08036v1}
	{http://arxiv.org/abs/1911.08036v1} \BibitemShut {NoStop}%
	\bibitem [{\citenamefont {Zangeneh}\ \emph {et~al.}(2019)\citenamefont
		{Zangeneh}, \citenamefont {Avdoshenko}, \citenamefont {van~den Brink},\ and\
		\citenamefont {Hozoi}}]{Zangeneh2019}%
	\BibitemOpen
	\bibfield  {author} {\bibinfo {author} {\bibfnamefont {Z.}~\bibnamefont
			{Zangeneh}}, \bibinfo {author} {\bibfnamefont {S.}~\bibnamefont
			{Avdoshenko}}, \bibinfo {author} {\bibfnamefont {J.}~\bibnamefont {van~den
				Brink}},\ and\ \bibinfo {author} {\bibfnamefont {L.}~\bibnamefont {Hozoi}},\
	}\bibfield  {title} {\bibinfo {title} {Single-site magnetic anisotropy
			governed by interlayer cation charge imbalance in triangular-lattice
			\ce{AYbX2}},\ }\href@noop {} {\bibfield  {journal} {\bibinfo  {journal}
			{Phys. Rev. B}\ }\textbf {\bibinfo {volume} {100}} (\bibinfo {year}
		{2019})}\BibitemShut {NoStop}%
	\bibitem [{\citenamefont {Güntherodt}\ \emph {et~al.}(1983)\citenamefont
		{Güntherodt}, \citenamefont {Jayaraman}, \citenamefont {Batlogg},
		\citenamefont {Croft},\ and\ \citenamefont {Melczer}}]{Guentherodt1983}%
	\BibitemOpen
	\bibfield  {author} {\bibinfo {author} {\bibfnamefont {G.}~\bibnamefont
			{Güntherodt}}, \bibinfo {author} {\bibfnamefont {A.}~\bibnamefont
			{Jayaraman}}, \bibinfo {author} {\bibfnamefont {G.}~\bibnamefont {Batlogg}},
		\bibinfo {author} {\bibfnamefont {M.}~\bibnamefont {Croft}},\ and\ \bibinfo
		{author} {\bibfnamefont {E.}~\bibnamefont {Melczer}},\ }\bibfield  {title}
	{\bibinfo {title} {Raman scattering from coupled phonon and electronic
			crystal-field excitations in \ce{CeAl2}},\ }\href@noop {} {\bibfield
		{journal} {\bibinfo  {journal} {Phys. Rev. Lett.}\ }\textbf {\bibinfo
			{volume} {51}},\ \bibinfo {pages} {2330} (\bibinfo {year}
		{1983})}\BibitemShut {NoStop}%
	\bibitem [{\citenamefont {Maczka}\ \emph {et~al.}(2008)\citenamefont
		{Maczka}, \citenamefont {Sanju{\'{a}}n}, \citenamefont {Fuentes},
		\citenamefont {Hermanowicz},\ and\ \citenamefont {Hanuza}}]{Maczka2008}%
	\BibitemOpen
	\bibfield  {author} {\bibinfo {author} {\bibfnamefont {M.}~\bibnamefont
			{Maczka}}, \bibinfo {author} {\bibfnamefont {M.~L.}\ \bibnamefont
			{Sanju{\'{a}}n}}, \bibinfo {author} {\bibfnamefont {A.~F.}\ \bibnamefont
			{Fuentes}}, \bibinfo {author} {\bibfnamefont {K.}~\bibnamefont
			{Hermanowicz}},\ and\ \bibinfo {author} {\bibfnamefont {J.}~\bibnamefont
			{Hanuza}},\ }\bibfield  {title} {\bibinfo {title} {Temperature-dependent
			raman study of the spin-liquid pyrochlore \ce{Tb2Ti2O7}},\ }\href@noop {}
	{\bibfield  {journal} {\bibinfo  {journal} {Phys. Rev. B}\ }\textbf {\bibinfo
			{volume} {78}} (\bibinfo {year} {2008})}\BibitemShut {NoStop}%
	\bibitem [{\citenamefont {Sethi}\ \emph {et~al.}(2019)\citenamefont {Sethi},
		\citenamefont {Slimak}, \citenamefont {Kolodiazhnyi},\ and\ \citenamefont
		{Cooper}}]{Sethi2019}%
	\BibitemOpen
	\bibfield  {author} {\bibinfo {author} {\bibfnamefont {A.}~\bibnamefont
			{Sethi}}, \bibinfo {author} {\bibfnamefont {J.}~\bibnamefont {Slimak}},
		\bibinfo {author} {\bibfnamefont {T.}~\bibnamefont {Kolodiazhnyi}},\ and\
		\bibinfo {author} {\bibfnamefont {S.}~\bibnamefont {Cooper}},\ }\bibfield
	{title} {\bibinfo {title} {Emergent vibronic excitations in the
			magnetodielectric regime of \ce{Ce2O3}},\ }\href@noop {} {\bibfield
		{journal} {\bibinfo  {journal} {Phys. Rev. Lett.}\ }\textbf {\bibinfo
			{volume} {122}} (\bibinfo {year} {2019})}\BibitemShut {NoStop}%
	\bibitem [{\citenamefont {Hense}\ \emph {et~al.}(2004)\citenamefont {Hense},
		\citenamefont {Gratz}, \citenamefont {Nowotny},\ and\ \citenamefont
		{Hoser}}]{Hense_2004}%
	\BibitemOpen
	\bibfield  {author} {\bibinfo {author} {\bibfnamefont {K.}~\bibnamefont
			{Hense}}, \bibinfo {author} {\bibfnamefont {E.}~\bibnamefont {Gratz}},
		\bibinfo {author} {\bibfnamefont {H.}~\bibnamefont {Nowotny}},\ and\ \bibinfo
		{author} {\bibfnamefont {A.}~\bibnamefont {Hoser}},\ }\bibfield  {title}
	{\bibinfo {title} {Lattice dynamics and the interaction with the crystal
			electric field in \ce{NdCu2}},\ }\href@noop {} {\bibfield  {journal}
		{\bibinfo  {journal} {J. Phys.: Condens. Matter}\ }\textbf {\bibinfo {volume}
			{16}},\ \bibinfo {pages} {5751} (\bibinfo {year} {2004})}\BibitemShut
	{NoStop}%
	\bibitem [{\citenamefont {Arnold}\ \emph {et~al.}(2014)\citenamefont {Arnold},
		\citenamefont {Bilheux}, \citenamefont {Borreguero}, \citenamefont {Buts},
		\citenamefont {Campbell}, \citenamefont {Chapon}, \citenamefont {Doucet},
		\citenamefont {Draper}, \citenamefont {Leal}, \citenamefont {Gigg},
		\citenamefont {Lynch}, \citenamefont {Markvardsen}, \citenamefont
		{Mikkelson}, \citenamefont {Mikkelson}, \citenamefont {Miller}, \citenamefont
		{Palmen}, \citenamefont {Parker}, \citenamefont {Passos}, \citenamefont
		{Perring}, \citenamefont {Peterson}, \citenamefont {Ren}, \citenamefont
		{Reuter}, \citenamefont {Savici}, \citenamefont {Taylor}, \citenamefont
		{Taylor}, \citenamefont {Tolchenov}, \citenamefont {Zhou},\ and\
		\citenamefont {Zikovsky}}]{Arnold2014}%
	\BibitemOpen
	\bibfield  {author} {\bibinfo {author} {\bibfnamefont {O.}~\bibnamefont
			{Arnold}}, \bibinfo {author} {\bibfnamefont {J.}~\bibnamefont {Bilheux}},
		\bibinfo {author} {\bibfnamefont {J.}~\bibnamefont {Borreguero}}, \bibinfo
		{author} {\bibfnamefont {A.}~\bibnamefont {Buts}}, \bibinfo {author}
		{\bibfnamefont {S.}~\bibnamefont {Campbell}}, \bibinfo {author}
		{\bibfnamefont {L.}~\bibnamefont {Chapon}}, \bibinfo {author} {\bibfnamefont
			{M.}~\bibnamefont {Doucet}}, \bibinfo {author} {\bibfnamefont
			{N.}~\bibnamefont {Draper}}, \bibinfo {author} {\bibfnamefont {R.~F.}\
			\bibnamefont {Leal}}, \bibinfo {author} {\bibfnamefont {M.}~\bibnamefont
			{Gigg}}, \bibinfo {author} {\bibfnamefont {V.}~\bibnamefont {Lynch}},
		\bibinfo {author} {\bibfnamefont {A.}~\bibnamefont {Markvardsen}}, \bibinfo
		{author} {\bibfnamefont {D.}~\bibnamefont {Mikkelson}}, \bibinfo {author}
		{\bibfnamefont {R.}~\bibnamefont {Mikkelson}}, \bibinfo {author}
		{\bibfnamefont {R.}~\bibnamefont {Miller}}, \bibinfo {author} {\bibfnamefont
			{K.}~\bibnamefont {Palmen}}, \bibinfo {author} {\bibfnamefont
			{P.}~\bibnamefont {Parker}}, \bibinfo {author} {\bibfnamefont
			{G.}~\bibnamefont {Passos}}, \bibinfo {author} {\bibfnamefont
			{T.}~\bibnamefont {Perring}}, \bibinfo {author} {\bibfnamefont
			{P.}~\bibnamefont {Peterson}}, \bibinfo {author} {\bibfnamefont
			{S.}~\bibnamefont {Ren}}, \bibinfo {author} {\bibfnamefont {M.}~\bibnamefont
			{Reuter}}, \bibinfo {author} {\bibfnamefont {A.}~\bibnamefont {Savici}},
		\bibinfo {author} {\bibfnamefont {J.}~\bibnamefont {Taylor}}, \bibinfo
		{author} {\bibfnamefont {R.}~\bibnamefont {Taylor}}, \bibinfo {author}
		{\bibfnamefont {R.}~\bibnamefont {Tolchenov}}, \bibinfo {author}
		{\bibfnamefont {W.}~\bibnamefont {Zhou}},\ and\ \bibinfo {author}
		{\bibfnamefont {J.}~\bibnamefont {Zikovsky}},\ }\bibfield  {title} {\bibinfo
		{title} {Mantid-data analysis and visualization package for neutron
			scattering and $\mu$sr experiments},\ }\href@noop {} {\bibfield  {journal}
		{\bibinfo  {journal} {Nucl. Instrum. Methods. Phys. Res. A}\ }\textbf
		{\bibinfo {volume} {764}},\ \bibinfo {pages} {156} (\bibinfo {year}
		{2014})}\BibitemShut {NoStop}%
	\bibitem [{\citenamefont {Andrew}\ \emph {et~al.}(2019)\citenamefont {Andrew},
		\citenamefont {Applin}, \citenamefont {Arnold}, \citenamefont {Bamidele},
		\citenamefont {Basso}, \citenamefont {Borreguero}, \citenamefont {Brown},
		\citenamefont {Brown}, \citenamefont {Draper}, \citenamefont {Ganeva},
		\citenamefont {Gigg}, \citenamefont {Guest}, \citenamefont {Hahn},
		\citenamefont {Heybrock}, \citenamefont {Jackson}, \citenamefont {Jenkins},
		\citenamefont {Jones}, \citenamefont {Le}, \citenamefont {Leal},
		\citenamefont {Lim}, \citenamefont {Lin}, \citenamefont {Lynch},
		\citenamefont {McDonnell}, \citenamefont {Miladinovic}, \citenamefont
		{Moore}, \citenamefont {Nixon}, \citenamefont {Oram}, \citenamefont
		{Peterson}, \citenamefont {Reimund}, \citenamefont {Russell}, \citenamefont
		{Saunders}, \citenamefont {Savici}, \citenamefont {Soininen}, \citenamefont
		{Sokolova}, \citenamefont {Sullivan}, \citenamefont {Tasev}, \citenamefont
		{Titcombe}, \citenamefont {Tolchenov}, \citenamefont {Vardanyan},
		\citenamefont {Whitfield},\ and\ \citenamefont {Zhou}}]{Andrew2019}%
	\BibitemOpen
	\bibfield  {author} {\bibinfo {author} {\bibfnamefont {M.}~\bibnamefont
			{Andrew}}, \bibinfo {author} {\bibfnamefont {R.}~\bibnamefont {Applin}},
		\bibinfo {author} {\bibfnamefont {O.}~\bibnamefont {Arnold}}, \bibinfo
		{author} {\bibfnamefont {A.}~\bibnamefont {Bamidele}}, \bibinfo {author}
		{\bibfnamefont {L.}~\bibnamefont {Basso}}, \bibinfo {author} {\bibfnamefont
			{J.}~\bibnamefont {Borreguero}}, \bibinfo {author} {\bibfnamefont
			{E.}~\bibnamefont {Brown}}, \bibinfo {author} {\bibfnamefont
			{H.}~\bibnamefont {Brown}}, \bibinfo {author} {\bibfnamefont
			{N.}~\bibnamefont {Draper}}, \bibinfo {author} {\bibfnamefont
			{M.}~\bibnamefont {Ganeva}}, \bibinfo {author} {\bibfnamefont {M.~A.}\
			\bibnamefont {Gigg}}, \bibinfo {author} {\bibfnamefont {G.}~\bibnamefont
			{Guest}}, \bibinfo {author} {\bibfnamefont {S.}~\bibnamefont {Hahn}},
		\bibinfo {author} {\bibfnamefont {S.}~\bibnamefont {Heybrock}}, \bibinfo
		{author} {\bibfnamefont {A.~J.}\ \bibnamefont {Jackson}}, \bibinfo {author}
		{\bibfnamefont {S.}~\bibnamefont {Jenkins}}, \bibinfo {author} {\bibfnamefont
			{S.}~\bibnamefont {Jones}}, \bibinfo {author} {\bibfnamefont
			{D.}~\bibnamefont {Le}}, \bibinfo {author} {\bibfnamefont {R.}~\bibnamefont
			{Leal}}, \bibinfo {author} {\bibfnamefont {A.}~\bibnamefont {Lim}}, \bibinfo
		{author} {\bibfnamefont {J.}~\bibnamefont {Lin}}, \bibinfo {author}
		{\bibfnamefont {V.}~\bibnamefont {Lynch}}, \bibinfo {author} {\bibfnamefont
			{M.}~\bibnamefont {McDonnell}}, \bibinfo {author} {\bibfnamefont
			{G.}~\bibnamefont {Miladinovic}}, \bibinfo {author} {\bibfnamefont
			{L.}~\bibnamefont {Moore}}, \bibinfo {author} {\bibfnamefont
			{D.}~\bibnamefont {Nixon}}, \bibinfo {author} {\bibfnamefont
			{E.}~\bibnamefont {Oram}}, \bibinfo {author} {\bibfnamefont {P.~F.}\
			\bibnamefont {Peterson}}, \bibinfo {author} {\bibfnamefont {V.}~\bibnamefont
			{Reimund}}, \bibinfo {author} {\bibfnamefont {A.}~\bibnamefont {Russell}},
		\bibinfo {author} {\bibfnamefont {H.}~\bibnamefont {Saunders}}, \bibinfo
		{author} {\bibfnamefont {A.}~\bibnamefont {Savici}}, \bibinfo {author}
		{\bibfnamefont {A.}~\bibnamefont {Soininen}}, \bibinfo {author}
		{\bibfnamefont {A.}~\bibnamefont {Sokolova}}, \bibinfo {author}
		{\bibfnamefont {B.}~\bibnamefont {Sullivan}}, \bibinfo {author}
		{\bibfnamefont {D.}~\bibnamefont {Tasev}}, \bibinfo {author} {\bibfnamefont
			{T.}~\bibnamefont {Titcombe}}, \bibinfo {author} {\bibfnamefont
			{R.}~\bibnamefont {Tolchenov}}, \bibinfo {author} {\bibfnamefont
			{G.}~\bibnamefont {Vardanyan}}, \bibinfo {author} {\bibfnamefont
			{R.}~\bibnamefont {Whitfield}},\ and\ \bibinfo {author} {\bibfnamefont
			{W.}~\bibnamefont {Zhou}},\ }\href@noop {} {\bibinfo {title} {Mantid 4.1.0:
			Manipulation and analysis toolkit for instrument data}} (\bibinfo {year}
	{2019})\BibitemShut {NoStop}%
\end{thebibliography}
\end{document}